\documentclass[
superscriptaddress,
 amsmath,amssymb,
 aps,
prb,two column
]{revtex4-1}

\usepackage{graphicx}
\usepackage{dcolumn}
\usepackage{bm}

\usepackage{graphicx,epsf,bm,bbm,color,amsmath,ulem,amssymb,float, subfigure}

\usepackage{color}
\usepackage{ textcomp }
\usepackage{ dsfont }
\usepackage{indentfirst}
\newcommand{\remove}[1]{}

\newcommand{\bi}{\begin{itemize}}
\newcommand{\ei}{\end{itemize}}
\newcommand{\be}{\begin{enumerate}}
\newcommand{\ee}{\end{enumerate}}

\newenvironment{dfn}{{\vspace*{1ex} \noindent \bf Definition }}{\vspace*{1ex}}

	\newcommand{\beq}{\begin{eqnarray}}
	\newcommand{\eeq}{\end{eqnarray}}

\begin{document}


\title{Effects of Defects in Superconducting Phase of Twisted Bilayer Graphene}

\author{Hui Yang}
\altaffiliation{HY and ZQG contributed equally to this work.}
\affiliation{International Center for Quantum Materials, School of Physics, Peking University, Beijing 100871, China
}

\author{Zhi-Qiang Gao}
\altaffiliation{HY and ZQG contributed equally to this work.}
\affiliation{International Center for Quantum Materials, School of Physics, Peking University, Beijing 100871, China
}

\author{Fa Wang}
\affiliation{International Center for Quantum Materials, School of Physics, Peking University, Beijing 100871, China
}
\affiliation{Collaborative Innovation Center of Quantum Matter, Beijing 100871, China
}

\date{\today}

\begin{abstract}
In this work the effects of defects in the superconducting phases of the twisted bilayer graphene (TBG) are investigated. A well-accepted low energy effective model and a non-magnetic impurity potential to mimic defects are employed. Different superconducting pairing symmetries, including $s$-wave, $(d+id)$-wave and $(p+ip)$-wave pairing, are considered. In single impurity case, the local density of states (DOS) are calculated for the pairing symmetries above. For different pairing symmetries the number and property of bound states induced by defects are different. In multi-impurity case, the phase diagrams are calculated in terms of effective gap and the strength and density of impurities. In unconventional superconducting phases, namely $(p+ip)$-wave and $(d+id)$-wave phases, superconductivity will be destroyed by impurities with strong strength or concentration. These results can in principle be detected in scanning tunnelling microscopy (STM) experiments, and therefore the pairing symmetry, at least whether the superconductivity is conventional or unconventional, may be determined.
\end{abstract}

\maketitle


\section{\label{sec:level1}Introduction}
Twisted bilayer graphene (TBG) has attracted much interest these days. The most striking property of TBG is that flat bands emerge at the magic angle $\theta=1.08^{\circ}$\cite{Bistritzer_2011}. Due to the inherent strong-correlation nature of the flat bands, an insulating phase discovered at the  filling of $n=2$ is argued to be a Mott insulator\cite{Cao_2018}. Around the insulator phase, superconducting phases\cite{Cao_unconventional_2018} were observed by doping slightly away from the insulator phase. Different theories giving rise to different pairing symmetries have been proposed to explain the superconducting phases\cite{Xu:2018aa,Po:2018aa,PhysRevX.8.041041,PhysRevB.99.094521,PhysRevB.98.235158,lian_twisted_2018}. However, the pairing symmetry of the order parameter in superconducting phase of the TBG system is still under debate\cite{2019arXiv190207410T}. A most heated debate is whether the superconductivity is unconventional or simply conventional, which is believed to determine whether the origin of the superconductivity in TBG is correlation physics or merely electron-phonon coupling. One experimental method to identify the pairing symmetry in TBG has been proposed in Ref\cite{PhysRevB.99.220507}, which subjects the TBG to an external magnetic field and strain. In this work another method is proposed to distinguish the pairing symmetry in TBG by studying the impurity induced bound states in the superconductor phases. Since this method only involves the low energy effective theory of the system, it can be also extended to other lattice systems with Moir\'{e} pattern or quasi-crystal systems, where tight-binding models are hard to develop and the mostly practicable models are simply low energy effective models.

\begin{figure}[htbp]
\centering
\includegraphics[width=8cm]{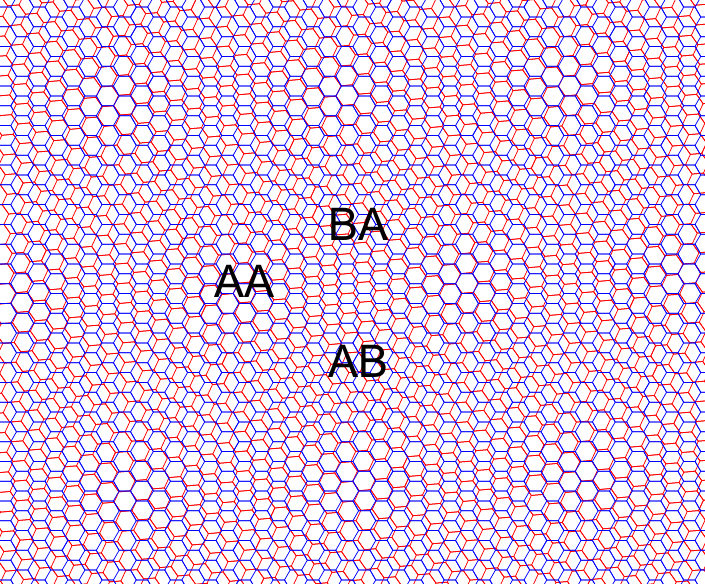}
\caption{The lattice of the TBG. $AA$, $AB$ and $BA$ regions are showed in the figure.}
\end{figure}

Impurities in superconductor may give rise to different phenomena for different pairing mechanism and different pairing symmetry\cite{RevModPhys.78.373}. A nonmagnetic impurity will not break the Cooper pair in an $s$-wave superconductor\cite{PhysRevLett.3.325}, but it can break Cooper pairs with $p$-wave and $d$-wave symmetry and may induce bound states or quasi-bound states inside the superconducting gap. A magnetic impurity may induce Kondo effect in the superconducting phase\cite{RevModPhys.78.373}. In multi-impurity case,when the strength and density of impurities is large, the superconducting phase coherence will be destroyed, which converts the system to a normal phase\cite{RevModPhys.78.373}. Since disorder such as carbon vacancy and adatom is unavoidable in graphene\cite{RevModPhys.81.109}, it is necessary to study the effect of impurities in TBG\cite{2019arXiv190802753W,2019arXiv190702856H,PhysRevB.99.245118,PhysRevMaterials.3.084003}.\\

We study the effect of impurity by calculating the number of impurity induced in-gap states for different pairing symmetry, from which we can get some knowledge about the pairing symmetry in the TBG system. However, the correlation between electrons are not considered, which is also believed to be important in TBG\cite{Cao_2018,Cao_unconventional_2018}. The in-gap states can be observed in STM experiments and may serve as an experimental indicator of the pairing symmetries, and the method proposed in this paper can be also employed in experiments on other systems with Moir\'{e} pattern. This paper is organized as follows. In Sec. II, the model proposed in Ref.\cite{Bistritzer_2011} is briefly reviewed and the BdG Hamiltonian is introduced to describe the superconductivity. In Sec. III, the single impurity effects in superconducting phases are investigated by calculating the DOS. We find that the number of bound state is different for different paring symmetries. In Sec. IV, the the multi-impurity effects are investigated in superconducting phases by calculating the effective superconducting gap as a function of the effective strength of impurities, which shows the extinction of superconductivity in $(d+id)$-wave phase and $(p+ip)$-wave phase. The conclusions are given in Sec. V. As a comparison, we employ a tight-binding model\cite{Liu:2018aa} as basis and study the impurity effects in that model in Appendix. 

\section{\label{sec:level1}The Model}

\subsection{The Model Describing Flat Bands}

The model proposed in Ref.\cite{Bistritzer_2011} is used to describe the flat band system without impurities.The Moir\'{e} bands Hamiltonian\cite{Bistritzer_2011} close to the Dirac point reads
\beq
H_{\vec{k}\vec{k}^{\prime}}=\delta_{\vec{k}\vec{k}^{\prime}}
\begin{pmatrix}
h_{\vec{k}}(\frac{\theta}{2}) & wT_{1} & wT_{2} & wT_{3}\\
wT_{1}^{\dagger} & h_{\vec{k}+\vec{q}_1}(-\frac{\theta}{2}) & 0 & 0\\
wT_{2}^{\dagger} & 0 & h_{\vec{k}+\vec{q}_2}(-\frac{\theta}{2}) & 0\\
wT_{3}^{\dagger} & 0 & 0 & h_{\vec{k}+\vec{q}_3}(-\frac{\theta}{2})\\
\end{pmatrix},\nonumber
\eeq
where $w=110$ meV is the strength of hopping and $\vec{q}_{j}$ are defined in FIG. 2(a).
\begin{figure}[htbp]
\centering
\includegraphics[width=8cm]{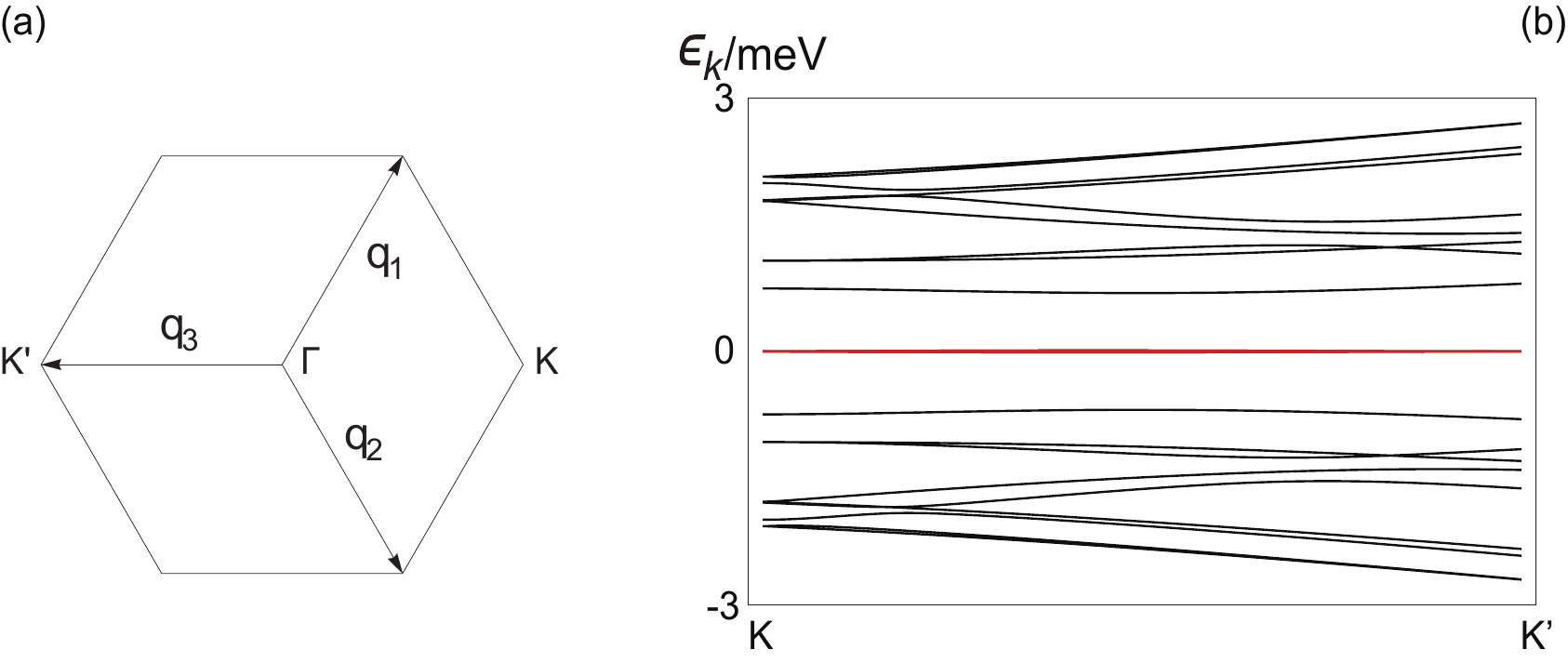}
\caption{(a) The three $\vec{q}_{j}$ defined in Moiré Brillouin zone. (b) Dispersion relations of Bistritzer-MacDonald model. Each flat band coloured as red has a 4-fold spin-valley degeneracy.}
\end{figure}
The matrix elements (which are all two by two matrices) of the Hamiltonian above are defined as\cite{Bistritzer_2011}
\beq
h_{\vec{k}}(\theta)&=&-vk
\begin{pmatrix}
0 & e^{i({\theta_{\vec{k}}}-\theta)}\\
e^{-i({\theta_{\vec{k}}}-\theta)} & 0\\
\end{pmatrix},
\eeq
\beq
T_{1}&=&
\begin{pmatrix}
1 & 1\\
1 & 1\\
\end{pmatrix},
\eeq
\beq
T_{2}&=&
\begin{pmatrix}
e^{-i\frac{2\pi}{3}} & 1\\
e^{i\frac{2\pi}{3}} & e^{-i\frac{2\pi}{3}}\\
\end{pmatrix},
\eeq
\beq
T_{3}&=&
\begin{pmatrix}
e^{i\frac{2\pi}{3}} & 1\\
e^{-i\frac{2\pi}{3}} & e^{i\frac{2\pi}{3}}\\
\end{pmatrix},
\eeq
where $h_{\vec{k}}(\theta)$ is the Hamiltonian of graphene and $v$ is the Dirac velocity. Besides, ${\vec{k}}$ is measured from Dirac points. Dispersion relations around Dirac point is shown in FIG. 2(b). Each flat band has a 4-fold spin-valley degeneracy.

The impurity potential in real space is $U_{\text{imp}}(\vec{r})=u\delta_{\vec{r},\vec{R}_0}$, where $u$ is the strength of the potential and $\vec{R}_{0}$ is the location of the impurity. The impurity potential is quantized and projected to the Hilbert space of the flat bands.

\subsection{\label{sec:level2}Pairing Symmetry}

To describe the superconductivity, the BdG Hamiltonian is introduced
\beq
H^{BdG}(\vec{k})=
\begin{pmatrix}
E(\vec{k})-\mu & -\Delta(\vec{k})\\
-\Delta^{\dagger}(\vec{k}) & -E(\vec{k})+\mu\\
\end{pmatrix},
\eeq
where diagonal matrix $E(\vec{k})$ represents the flat bands in spin space, $\mu$ is the chemical potential, and $\Delta(\vec{k})$ is the order parameter matrix. Since there is no interband pairing\cite{Liu:2018aa}, $\Delta(\vec{k})$ is diagonal and then gains the form of $\Delta(\vec{k})=S(\vec{k})\cdot \text{diag}\{\Delta_1,\Delta_2,\Delta_3,\Delta_4\}$, where $\Delta_{j}$ is the order parameter in the $j$-th band and  $S(\vec{k})$ reflects the symmetry of them. For $s$-wave, $S(\vec{k})=1$. For ($d+id$)-wave\cite{Liu:2018aa}  $S(\vec{k})=k_1^2-k_2^2+ik_1k_2$. For  ($p+ip$)-wave\cite{Liu:2018aa}, $S(\vec{k})=k_1+ik_2$. The value of $k_1$ and $k_2$ are defined as the projection of $\vec{k}$ on the direction of $\vec{q}_1$ and $\vec{q}_2$, respectively.

\section{Bound States Induced by Impurities in Superconducting Phase of TBG}

\begin{figure*}[htbp]
\centering
\includegraphics[width=16cm]{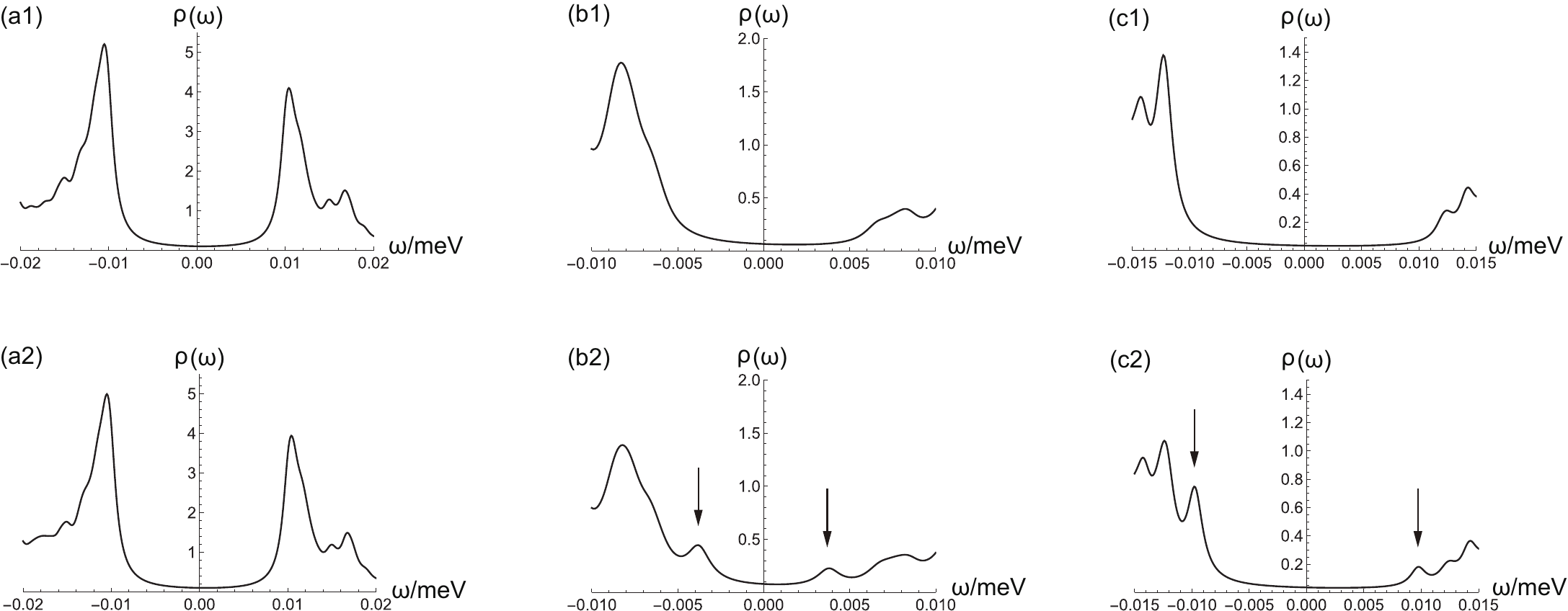}
\caption{The local DOS. (a1) and (a2) The local DOS for $s$-wave phase. (b1) and (b2) The local DOS for $(d+id)$-wave phase. (c1) and (c2) The local DOS for $(p+ip)$-wave phase. All bound states pointed by arrows are 2-fold degenerate.}
\end{figure*}

\begin{figure*}[htbp]
\centering
\includegraphics[width=16cm]{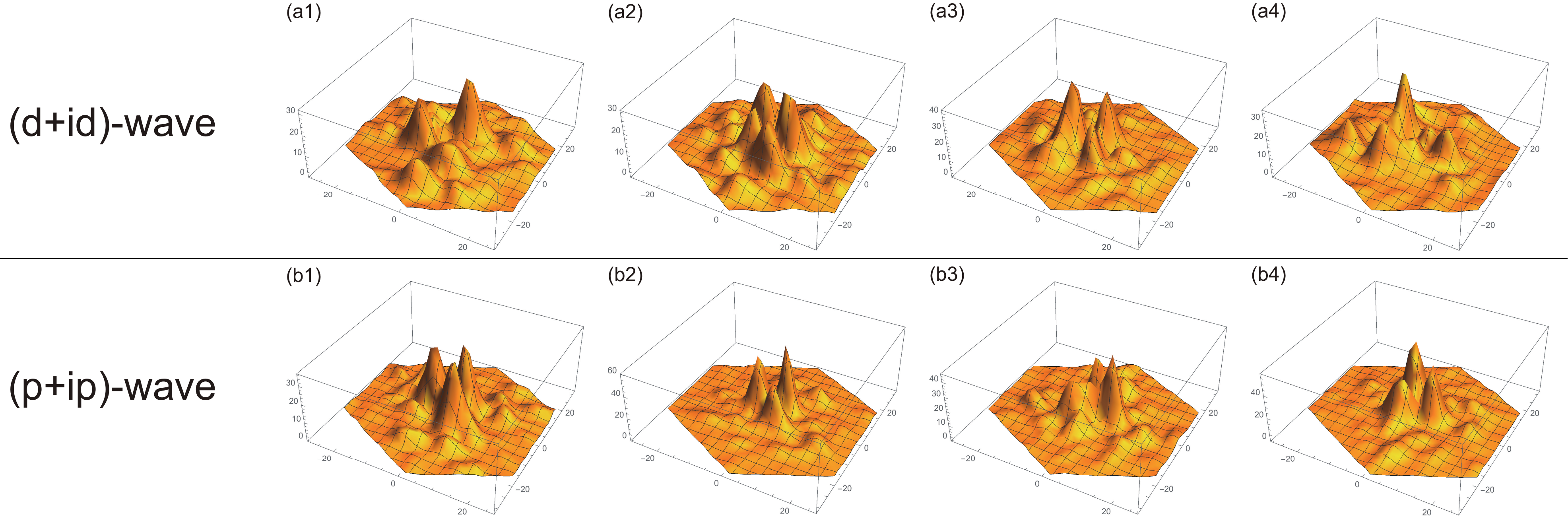}
\caption{The spatial distribution of the bound state wave functions. (a1) to (a4) The spatial distribution of the bound state wave functions for $(d+id)$-wave phase. (b1) to (b4) The spatial distribution of the bound state wave functions for $(p+ip)$-wave phase.}
\end{figure*}

After the impurity potential is projected to the Hilbert space of flat bands, the local DOS can be calculated by T-Matrix method\cite{RevModPhys.78.373}. Results are shown in FIG. 3. Due to the restriction of numerical calculation resource, the Lorentz broadening has to be enlarged to smooth the curves, which inevitable leads to blunt peaks. The coefficients are set $u=1.0$ meV and $\Delta_{1}=\Delta_{2}=\Delta_{3}=\Delta_{4}=0.01$ meV for $s$-wave phase, $u=0.0001$ meV and $\Delta_{1}=\Delta_{2}=\Delta_{3}=\Delta_{4}=0.1$ meV for $(d+id)$-wave phase, and $u=1.0$ meV and $\Delta_{1}=\Delta_{2}=\Delta_{3}=\Delta_{4}=0.03$ meV for $(p+ip)$-wave phase. The chemical potential is set as $\mu=0.015$ meV to tune the filling around the electron half-filling. The in-gap states are identified as bound states. For $(d+id)$-wave phase, the $u$ is set to be much smaller because only when $u$ is small do the bound states emerge. This implies that when $u$ is large, the bound states in $(d+id)$-wave phase lies very close to or outside the band edge and thus hard to identify. The spatial distribution of the wave functions of these bound states shown in FIG. 4 shows that these bound states are indeed bounded around the impurity in real space.

From the local DOS, we found that bound states only emerge in unconventional phases, namely $(d+id)$-wave and $(p+ip)$-wave phase, and in $(d+id)$-wave phase only impurities with weak strength can induce observable bound states. The differences in number and property of bound states can be an effective tool to reveal the pairing symmetry in these  superconducting phases. It can serve as an indicator to determine pairing symmetry of the order parameter in the superconducting phase of the TBG.

Besides, each bound state shown in the local DOS is actually 2-fold degenerate. This degeneracy can be explained by the Kramers theorem. When the superconducting order parameter $\Delta(\vec{k})=S(\vec{k})\cdot \text{diag}\{\Delta_1,\Delta_2,\Delta_3,\Delta_4\}$ satisfies $\Delta_1=\Delta_3$ and $\Delta_2=\Delta_4$, a time-reversal-like symmetry $\mathcal{S}=i\tau^y \mathcal{K}$, where $\tau^y$ is the Pauli matrix in valley space, will protect the degeneracy. When this constraint of superconducting order parameter is broken, the 2-fold degeneracy will be consequently lifted. Further numerical results confirm this explanation.

\section{Phase Diagrams}

In this section the disorder average is applied to determine the phase diagrams which reflect how the effective superconducting gap relies on the density and strength of impurities $(na^2)^2 u$, where $n$ is the density of impurities, $a$ is the lattice constant and $u$ is the average strength of the impurity. Keeping other coefficients invariant, $(na^2)^2 u$ is varied from 0 to 0.00015 meV and the corresponding values of the effective gap are identified. Results are shown in FIG. 5.

\begin{figure}[htbp]
\centering
\includegraphics[width=8cm]{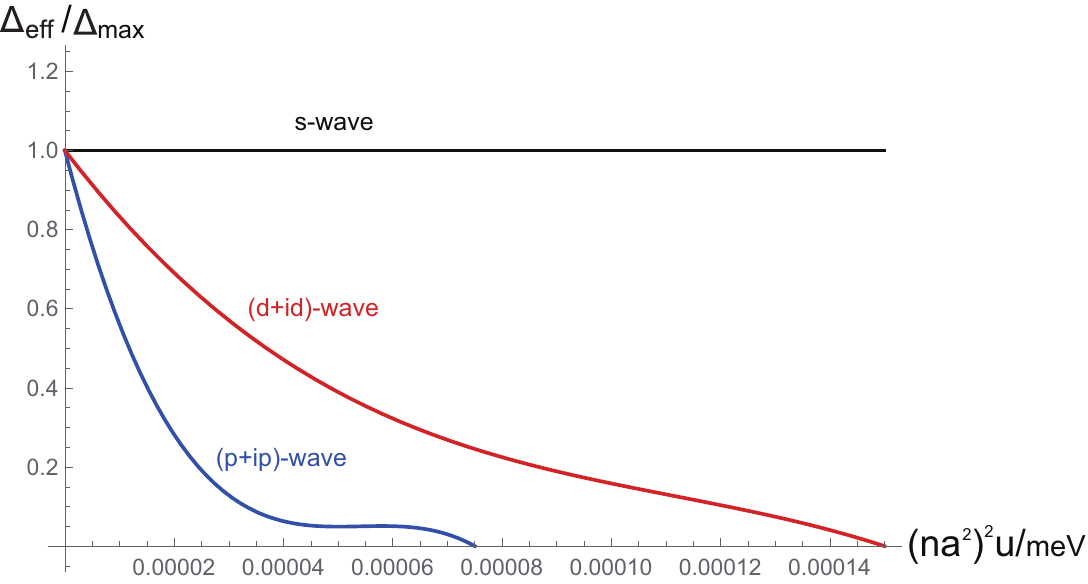}
\caption{Phase diagrams. Enough strength or concentration of impurities will destroy $(d+id)$-wave and $(p+ip)$-wave superconductivity, while the superconducting gap of $s$-wave phases can remain finite.}
\end{figure}

According to the phase diagrams, effects of impurities in different superconductivity phases are different. In unconventional $(d+id)$-wave phase and $(p+ip)$-wave phase, strong or dense impurities will destroy the superconductivity while in conventional $s$-wave phase they will not.

\section{\label{sec:level1}Conclusion}

In single impurity case, we find that for different pairing symmetries, the number and property of bound states are different. The results can be summarized in the table below.
\begin{table}[!htbp]
\begin{tabular}{|c|c|c|c|}
\hline
Paring Symmetry & \quad $s$  \quad & $d+id$ & $p+ip$\\
\hline
Number of Bound States & 0 & 2* & 2\\
\hline
\end{tabular}
\centering
\caption{Number of bound states for different pairing symmetries. The "*" means that bound states only emerge when the impurity strength is small.}
\end{table}
For $s$-wave phase, bound states never emerge; for $(d+id)$-wave phase, bound states emerge only when the impurity strength $u$ is small; and for $(p+ip)$-wave phase bound states always emerge. Thus, the number and property of bound states can serve as an indicator to show in which superconducting phase the TBG system is. In multi-impurity case, when the strength or density of impurities is large, superconductivity in $(d+id)$-wave phase and $(p+ip)$-wave phase can be fully destroyed.

In both the single impurity case and the multi-impurity case, the conventional and unconventional superconducting phases behave significantly different in their responses to impurities. Conventional superconducting phase has no bound state induced by a single defect, and is immune to strong or dense impurities, while unconventional superconducting phases have bound states and are not immune. Therefore, the defferent responses of different superconducting phases to impurities can be helpful to determine whether the superconductivity in the TBG is conventional or unconventional in experiments. This may answer whether the superconductivity in TBG is induced by correlated physics or traditional electron-phonon coupling, and hence improve the understanding to the TBG. Recently, STM experiments have successfully detected the local DOS of the TBG\cite{Xie2019,Kerelsky2019,Jiang2019,2019arXiv190102997C} without impurities. There are some methods to introduce defects into the graphene\cite{doi:10.1063/1.4945587,gonzalez-herrero_atomic-scale_2016}, and the effect of defects can be then detected by STM experiments. We hope further STM results can determine the pairing symmetry of the superconducting TBG by examining the number of bound states. The method proposed in this work only relies on the low energy effective model of the system, and insensitive to the high energy details. Therefore, it can be also employed in other Moir\'{e} systems, even quasi-crystal systems, where tight-binding models are hard to develop.

\section{\label{sec:level1}Acknowledgements}

ZQG thanks Congjun Wu, Yi-Zhuang You, Kai-Wei Sun and Ji-Chen Feng for helpful discussions. FW acknowledges support from The National Key Research and Development Program of China (Grand No. 2017YFA0302904).

\appendix

\section{The Results Based on a Tight-Binding Model}

Some previous works\cite{Yuan:2018aa,Po:2018aa,Xu:2018aa,Liu:2018aa} have proposed different models for the TBG. Among these models, we choose the four-band tight-binding model proposed by Ref\cite{Yuan:2018aa}. In this model, the BdG Hamiltonian reads
\beq
H^{BdG}_{\vec{k}}=
\begin{pmatrix}
E_{\vec{k}}-\mu & -\Delta_{\vec{k}}\\
-\Delta_{\vec{k}}^{\dagger} & -E_{\vec{k}}+\mu\\
\end{pmatrix},
\eeq
where diagonal matrix $E(\vec{k})$ represents the flat bands in spin space, $\mu$ is the chemical potential, and $\Delta_{\vec{k}}$ is the order parameter matrix. Since there is no interband pairing\cite{Liu:2018aa}, $\Delta_{\vec{k}}$ is diagonal and then gains the form of $\Delta_{\vec{k}}=S(\vec{k})\cdot \text{diag}\{\Delta_1,\Delta_2,\Delta_3,\Delta_4\}$, where $\Delta_{j}$ is the order parameter in the $j$-th band and  $S(\vec{k})$ reflects the symmetry of them. For $s$-wave, $S(\vec{k})=1$. For $d$-wave\cite{Liu:2018aa} ($d+id$), $S(\vec{k})=\cos(\frac{\sqrt{3}}{2} k_{x}+\frac{1}{2}k_{y})-\cos(\frac{\sqrt{3}}{2} k_{x}-\frac{1}{2}k_{y})+i\cdot \sin(\frac{\sqrt{3}}{2} k_{x}+\frac{1}{2}k_{y})\sin(\frac{\sqrt{3}}{2} k_{x}-\frac{1}{2}k_{y})$. For $p$-wave\cite{Liu:2018aa} ($p+ip$), $S(\vec{k})=\sin(\frac{\sqrt{3}}{2} k_{x}+\frac{1}{2}k_{y})+i\cdot \sin(\frac{\sqrt{3}}{2} k_{x}-\frac{1}{2}k_{y})$. The values of $k_{x}$ and $k_{y}$ are measured in the unit of $2\pi/a$ where $a$ is the lattice constant of the TBG supercell lattice.

This model reduces the complicated TBG structure to a honeycomb lattice formed by $AB$ and $BA$ sites of the supercell of TBG. However, impurities can be anywhere in the TBG, not only on the $AB$ and $BA$ sites. For simplicity, we consider those impurities located on $AA$, $AB$ and $BA$ sites.

\subsection{Construction of  Impurity Hamiltonian}

First, we consider a single impurity located on an $AB$ site. In Bloch representation, the impurity Hamiltonian takes the form
\beq
\hat{H}_{imp}^{AB}=\sum_{\vec{k},\vec{k^{\prime}}}P_{\vec{k}}^{\dagger}\cdot \mathcal{U}_{\vec{k}\vec{k^{\prime}}}\cdot P_{\vec{k^{\prime}}},
\eeq
where
\beq
P_{\vec{k}}=
\begin{pmatrix}
p_{x,\vec{k}}^{A} & p_{y,\vec{k}}^{A} & p_{x,\vec{k}}^{B} & p_{y,\vec{k}}^{B}
\end{pmatrix}^\mathbf{T}
\eeq
stands for annihilation operators in Bloch basis and $\mathcal{U}_{\vec{k}\vec{k^{\prime}}}$ is a four by four matrix whose elements are overlaps of Bloch wave functions and the impurity potential. Converting the expression of impurity Hamiltonian to Wannier representation, we have
\beq
\hat{H}_{imp}^{AB}=\sum_{\vec{k},\vec{k^{\prime}}}W_{\vec{k}}^{\dagger}\cdot (H^{AB}_{imp})_{\vec{k}\vec{k^{\prime}}}\cdot W_{\vec{k^{\prime}}},
\eeq
where $W_{\vec{k}}$ is the annihilation operators in Wannier basis and $(H^{AB}_{imp})_{\vec{k}\vec{k^{\prime}}}$ takes the form (take its (1,2)-component as an example)
\beq
((H^{AB}_{imp})_{\vec{k}\vec{k^{\prime}}})_{(1,2)}=u_0\sum_{i,j}e^{-i(\vec{k}\cdot \vec{R}_{AB}^{i}-\vec{k}^{\prime}\cdot \vec{R}_{AB}^{j})}\cdot\nonumber\\
w_{x}^{AB*}(\vec{R}_{0}^{AB}-\vec{R}_{AB}^{i})w_{y}^{AB}(\vec{R}_{0}^{AB}-\vec{R}_{AB}^{j}),\label{eq:101}
\eeq
where $w_{\nu}^{s}(\vec{r}-\vec{R}_{AB}^{j})$, $\nu=x,y$ and $s=AB,BA$ are Wannier wave functions, and $\vec{R}_{0}^{AB}$ is the location of the impurity on an $AB$ site.

As indicated in Ref\cite{Po:2018aa}, the Wannier functions are localized in $AB$ and $BA$ region. Therefore, the contribution of $w_{\nu}^{s*}(\vec{R}_{0}^{AB}-\vec{R}_{AB}^{i})w_{\nu^{\prime}}^{s^{\prime}}(\vec{R}_{0}^{AB}-\vec{R}_{AB}^{j})$ is dominant only when $\vec{R}_{AB}^{i}$ and $\vec{R}_{AB}^{j}$ are both close to $\vec{R}_{0}^{AB}$. The term $w_{\nu}^{s*}(0)w_{\nu^{\prime}}^{s^{\prime}}(0)$ is about one order larger than the terms $w_{\nu}^{s*}(\vec{r}_{i})w_{\nu^{\prime}}^{s^{\prime}}(0)$ and $w_{\nu}^{s*}(0)w_{\nu^{\prime}}^{s^{\prime}}(\vec{r}_{i})$, while the latter two are one order larger than $w_{\nu}^{s*}(\vec{r}_{i})w_{\nu^{\prime}}^{s^{\prime}}(\vec{r}_{j})$, $i,j=1,2,3$. We only include those terms above.
\begin{figure}[htbp]
\centering
\includegraphics[width=3cm]{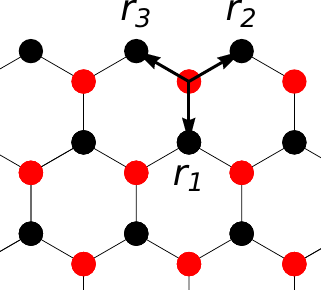}
\caption{The three $\vec{r}_{i}$ defined in the super lattice of the TBG.}
\end{figure}

As a result, in Eq. \ref{eq:101}, we only need to take account of terms that for both $i$ and $j$, $\vec{R}_{0}^{AB}-\vec{R}_{AB}^{i,j}$ equals to 0 or $\vec{r}_{l}$, $l=1,2,3$. With this preparation, we can construct our Hamiltonian for impurities located on $AB$ sites as
\beq
(H^{AB}_{imp})_{\vec{k}\vec{k^{\prime}}}=u\sum_{\vec{R}^{AB}_{imp}}e^{-i(\vec{k}-\vec{k^{\prime}})\cdot \vec{R}^{AB}_{imp}}\nonumber\\
\cdot \begin{pmatrix}
t_0\cdot \mathds{I}_{2\times 2} & T_{NN}\cdot J_{\vec{k}\vec{k}^{\prime}}\\
T_{NN}^{\dagger}\cdot J_{\vec{k}\vec{k}^{\prime}}^{*}& 0_{2\times 2}
\end{pmatrix},
\eeq
where we absorb the unit of energy into $t_0$ and $T_{NN}$, and left a dimensionless scaling factor $u$ to reflect the strength of the impurity. $\vec{R}^{AB}_{imp}$ is the position of impurities and $J_{\vec{k}\vec{k}^{\prime}}=\sum_{j=1,2,3}e^{-i(\vec{k}-\vec{k^{\prime}})\cdot \vec{r}_{j}}$. The value of coefficient $t_0$ which matches $w_{\nu}^{s*}(0)w_{\nu^{\prime}}^{s^{\prime}}(0)$, is about one order larger than the components of two by two matrix $T_{NN}$ which match $w_{\nu}^{s*}(\vec{r}_{i})w_{\nu^{\prime}}^{s^{\prime}}(0)$ and $w_{\nu}^{s*}(0)w_{\nu^{\prime}}^{s^{\prime}}(\vec{r}_{i})$. Besides, since the impurity Hamiltonian should conserve the point group symmetry of and time reversal symmetry, there are some restrictions on matrix $T_{NN}$. Given that the Wannier basis forms a four-dimensional representations of the point group of the TBG and the corresponding representation matrix of $C_{3}$ rotation is
\beq
C_{3}=
\begin{pmatrix}
\cos\theta & \sin\theta & 0 & 0\\
-\sin\theta & \cos\theta & 0 & 0\\
0 & 0 & \cos\theta & \sin\theta\\
0 & 0 & -\sin\theta & \cos\theta
\end{pmatrix},
\eeq
where $\theta=\frac{2\pi}{3}$. Then the impurity Hamiltonian should satisfy
\beq
C_{3}^{-1} \cdot (H^{AB}_{imp})_{\vec{k}\vec{k^{\prime}}} \cdot C_{3}=(H^{AB}_{imp})_{\vec{k}\vec{k^{\prime}}},
\eeq
which gives $T_{NN}$ the form
\beq
T_{NN}=
\begin{pmatrix}
t_{NN} & t_{NN}^\prime\\
-t_{NN}^\prime & t_{NN}
\end{pmatrix}.
\eeq
When superconductivity is taken accounted, the time reversal symmetry for the impurity Hamiltonian as well as the property of Hermitian requires that 
\beq
(H^{AB}_{imp})_{\vec{k}\vec{k^{\prime}}}=((H^{AB}_{imp})_{\vec{k^{\prime}}\vec{k}})^{\dagger}=((H^{AB}_{imp})_{-\vec{k^{\prime}},-\vec{k}})^{\mathbf{T}},
\eeq
which further indicates that $T_{NN}$ must be a real matrix.

Swapping the two columns and two rows of $(H^{AB}_{imp})_{\vec{k}\vec{k^{\prime}}}$, we can get the Hamiltonian for impurities located on $BA$ sites
\beq
(H^{BA}_{imp})_{\vec{k}\vec{k^{\prime}}}=u\sum_{\vec{R}^{BA}_{imp}}e^{-i(\vec{k}-\vec{k^{\prime}})\cdot \vec{R}^{BA}_{imp}}\nonumber\\
\cdot \begin{pmatrix}
0_{2\times 2} & T_{NN}^{\mathbf{T}}\cdot J_{\vec{k}\vec{k}^{\prime}}^{*}\\
T_{NN}\cdot J_{\vec{k}\vec{k}^{\prime}}& t_0\cdot \mathds{I}_{2\times 2}
\end{pmatrix},
\eeq
where we have already used $T_{NN}^{\dagger}=T_{NN}^{\mathbf{T}}$.

By the same argument, we can also construct the impurity Hamiltonian of $AA$ sites which only including terms of the same order of next-nearest-neighbour hopping
\beq
(H^{AA}_{imp})_{\vec{k}\vec{k^{\prime}}}=u\sum_{\vec{R}^{AA}_{imp}}e^{-i(\vec{k}-\vec{k^{\prime}})\cdot \vec{R}^{AA}_{imp}}\qquad\\
\cdot \begin{pmatrix}
t_{NNN}^{A}\cdot \mathds{I}_{2\times 2}\cdot J_{\vec{k}0}\cdot J_{0\vec{k}^{\prime}} & T_{NNN}\cdot J_{\vec{k}0}\cdot J_{\vec{k}^{\prime}0}\\
T_{NNN}^{\mathbf{T}}\cdot J_{0\vec{k}}\cdot J_{0\vec{k}^{\prime}} & t_{NNN}^{B}\cdot \mathds{I}_{2\times 2}\cdot J_{0,\vec{k}}\cdot J_{\vec{k}^{\prime},0}\nonumber
\end{pmatrix},
\eeq
where
\beq
T_{NNN}=
\begin{pmatrix}
t_{NNN} & t_{NNN}^\prime\\
-t_{NNN}^\prime & t_{NNN}
\end{pmatrix},
\eeq
and $t_{NNN}^{A}$, $t_{NNN}^{B}$, $t_{NNN}$ and $t_{NNN}^{\prime}$ are real coefficients of the same order as next-nearest-neighbour hoppings whose value are around 0.1 meV\cite{Liu:2018aa}.

\begin{figure}[htbp]
\centering
\includegraphics[width=8cm]{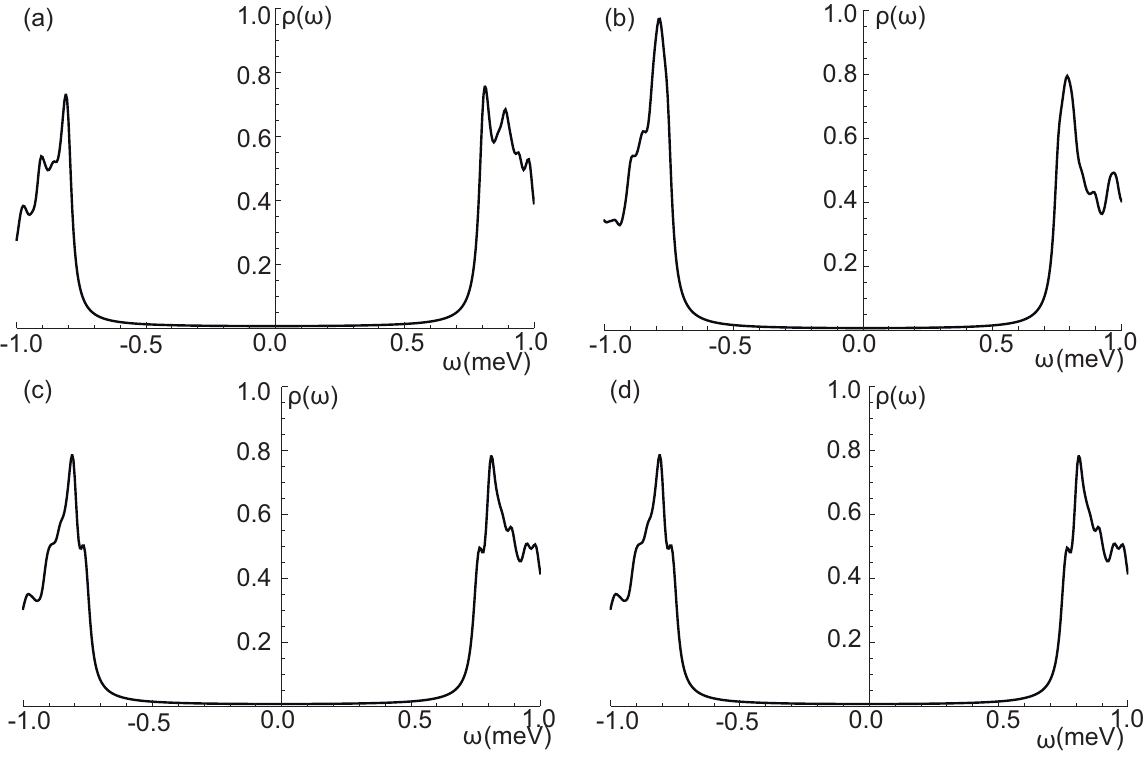}
\caption{Local DOS of $s$-wave phase. (a) shows the DOS without impurity, (b) shows the local DOS at the location of the impurity which is located at $AB$ or $BA$ region, (c) shows the local DOS at the nearest $AB$ region from the $AA$ region where the impurity is located and (d) shows the local DOS at the nearest $BA$ region from the $AA$ region where the impurity is located, respectively. For $s$-wave phase there is no bound state in the gap whenever the impurity is located at $AB$, $BA$ or $AA$ region.}
\end{figure}

\begin{figure}[htbp]
\centering
\includegraphics[width=8cm]{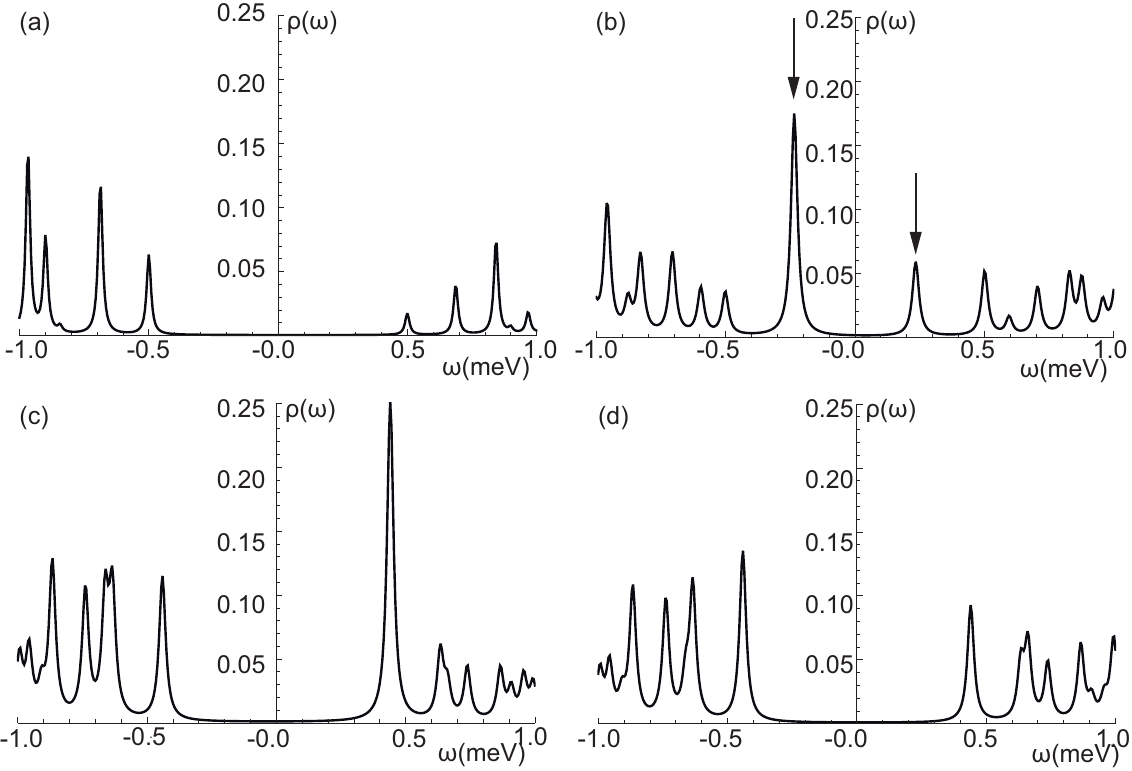}
\caption{Local DOS of $(d+id)$-wave phase. (a) The DOS without impurity. (b) The local DOS at the location of the impurity which is located at $AB$ or $BA$ region. (c) The local DOS at the nearest $AB$ region from the $AA$ region where the impurity is located. (d) The local DOS at the nearest $BA$ region from the $AA$ region where the impurity is located. For $(d+id)$-wave phase, there are two bound states in the gap when the impurity is located at $AB$ or $BA$ region and no bound state in the gap when the impurity is located at $AA$ region. Further numerical results corroborate that these bound states are indeed bounded around the impurity.}
\end{figure}

\begin{figure}[htbp]
\centering
\includegraphics[width=8cm]{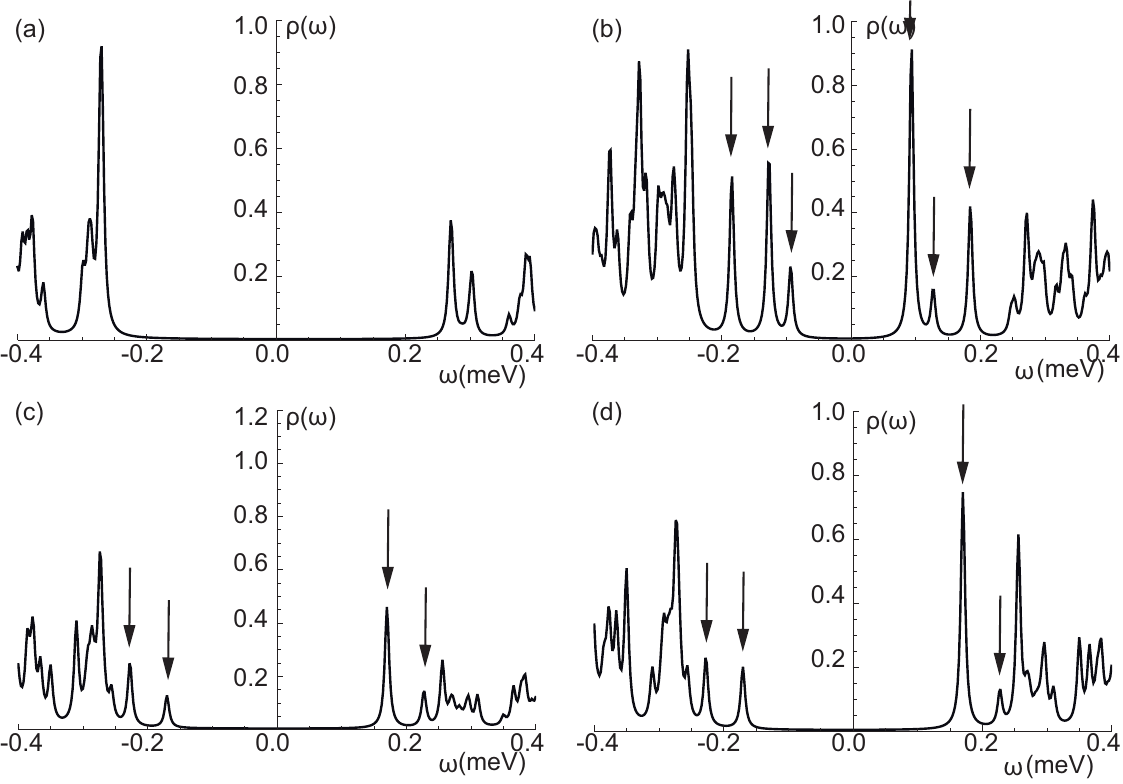}
\caption{Local DOS of $(p+ip)$-wave phase. (a) The DOS without impurity. (b) The local DOS at the location of the impurity which is located at $AB$ or $BA$ region. (c) The local DOS at the nearest $AB$ region from the $AA$ region where the impurity is located. (d) The local DOS at the nearest $BA$ region from the $AA$ region where the impurity is located. For $(p+ip)$-wave phase there are six bound states in the gap when the impurity is located at $AB$ or $BA$ region and four bound states in the gap when the impurity is located at $AA$ region. Further numerical results corroborate that these bound states are indeed bounded around the impurity.}
\end{figure}

\subsection{\label{sec:level2}Single Impurity and Local Density of State}

With preparation above, we can now calculate the local DOS by T-matrix method\cite{RevModPhys.78.373}. The local DOS for $s$-wave phase, $(d+id)$-wave phase and $(p+ip)$-wave phase are shown in FIG 7, 8 and 9, respectively. We set $\mu=-0.165$ meV. $\Delta_{1}=\Delta_{2}=0.6$ meV and $\Delta_{3}=\Delta_{4}=0.8$ meV for $s$-wave phase, $\Delta_{1}=\Delta_{2}=\Delta_{3}=\Delta_{4}=7.5$ meV for $(d+id)$-wave phase, and $\Delta_{1}=\Delta_{2}=0.8$ meV and $\Delta_{3}=\Delta_{4}=1.0$ meV for $(p+ip)$-wave phase. Other coefficients are set that $t_0=10.0$ meV, $t_{NN}=1.5$ meV, $t_{NN}^\prime=1.0$ meV, $t_{NNN}=0.2$ meV, $t_{NNN}^\prime=0.1$ meV, $t_{NNN}^{A}=0.2$ meV, $t_{NNN}^{B}=0.1$ meV,  $u=5.0$, $\Delta_{1}=\Delta_{2}=0.6$ meV and $\Delta_{3}=\Delta_{4}=0.8$ meV for $s$-wave phase, $\Delta_{1}=\Delta_{2}=\Delta_{3}=\Delta_{4}=7.5$ meV for $(d+id)$-wave phase, and $\Delta_{1}=\Delta_{2}=0.8$ meV and $\Delta_{3}=\Delta_{4}=1.0$ meV for $(p+ip)$-wave phase. The superconducting gaps we choose are larger than those observed in experiments; however, because of the restriction of computation resource, we have to enlarge these values to make our results numerically reliable. On the contrary to the results in the continuous model, the number of bound states is invariant when the strength of the impurity, which is represented by $u$, varies from 0.1 to 50. The results can be summarized in the table below.
\begin{table}[!htbp]
\begin{tabular}{|c|c|c|c|}
\hline
Impurity Location & \quad $s$  \quad & $d+id$ & $p+ip$\\
\hline
$AB$ region & 0 & 2 & 6\\
\hline
$AA$ region & 0 & 0 & 4\\
\hline
\end{tabular}
\centering
\caption{Number of bound states for different impurity locations and different kinds of pairing symmetry}
\end{table}

\subsection{\label{sec:level2}Phase Diagrams}

In this section we apply disorder average to determine the phase diagrams. Combining the BdG Hamiltonian and the impurity Hamiltonian, we arrive at
\beq
\hat{H}&=&\sum_{\vec{k},\vec{k}^{\prime}}\Psi_{\vec{k}}^{\dagger}\cdot (H^{BdG}_{\vec{k}}\delta_{\vec{k}\vec{k}^{\prime}}
+\sum_{\vec{R}^{AB}_{imp}}e^{-i(\vec{k}-\vec{k^{\prime}})\cdot \vec{R}^{AB}_{imp}}V_{\vec{k}\vec{k^{\prime}}}^{AB}\\
&+&\sum_{\vec{R}^{BA}_{imp}}e^{-i(\vec{k}-\vec{k^{\prime}})\cdot \vec{R}^{BA}_{imp}}V_{\vec{k}\vec{k^{\prime}}}^{BA}
+\sum_{\vec{R}^{AA}_{imp}}e^{-i(\vec{k}-\vec{k^{\prime}})\cdot \vec{R}^{AA}_{imp}}V_{\vec{k}\vec{k^{\prime}}}^{AA})\cdot \Psi_{\vec{k}}\nonumber ,
\eeq
where $\Psi_{\vec{k}}$ is the Nambu spinor and impurity scattering vertices $V^{\text{site}}_{\vec{k}\vec{k}^{\prime}}$s are defined as
\beq
V_{\vec{k}\vec{k^{\prime}}}^{\text{site}}&=&
\begin{pmatrix}
U_{\vec{k}\vec{k^{\prime}}}^{\text{site}} & 0\\
0 & -(U_{\vec{k^{\prime}}\vec{k}}^\text{site})^{\mathbf{T}}\\
\end{pmatrix},
\eeq
with
\beq
U_{\vec{k}\vec{k^{\prime}}}^{AB}&=&C_{\vec{k}}\cdot
u\begin{pmatrix}
a\cdot \mathds{I}_{2\times 2} & T_{NN}\cdot J_{\vec{k}\vec{k}^{\prime}}\\
T_{NN}^{\dagger}\cdot J_{\vec{k}\vec{k}^{\prime}}^{*}& 0_{2\times 2}\\
\end{pmatrix}
\cdot C_{\vec{k}^{\prime}}^{-1},\\
U_{\vec{k}\vec{k^{\prime}}}^{BA}&=&C_{\vec{k}}\cdot
u\begin{pmatrix}
0_{2\times 2} & T_{NN}^{\mathbf{T}}\cdot J_{\vec{k}\vec{k}^{\prime}}^{*}\\
T_{NN}\cdot J_{\vec{k}\vec{k}^{\prime}}& a\cdot \mathds{I}_{2\times 2}\\
\end{pmatrix}
\cdot C_{\vec{k}^{\prime}}^{-1},\\
U_{\vec{k}\vec{k^{\prime}}}^{AA}&=&C_{\vec{k}}\cdot
M_{\vec{k}\vec{k}^{\prime}}
\cdot C_{\vec{k}^{\prime}}^{-1},
\eeq
where
\beq
M_{\vec{k}\vec{k}^{\prime}}=u\begin{pmatrix}
t_{NNN}^{A}\cdot \mathds{I}_{2\times 2}\cdot J_{\vec{k}0}\cdot J_{0\vec{k}^{\prime}} & T_{NNN}\cdot J_{\vec{k}0}\cdot J_{\vec{k}^{\prime}0}\\
T_{NNN}^{\mathbf{T}}\cdot J_{0\vec{k}}\cdot J_{0\vec{k}^{\prime}} & t_{NNN}^{B}\cdot \mathds{I}_{2\times 2}\cdot J_{0,\vec{k}}\cdot J_{\vec{k}^{\prime},0}\\
\end{pmatrix},\nonumber
\eeq
and $C_{\vec{k}}$ is the transform matrix between Wannier basis and diagonal basis.

\begin{figure}[htbp]
\centering
\includegraphics[width=8cm]{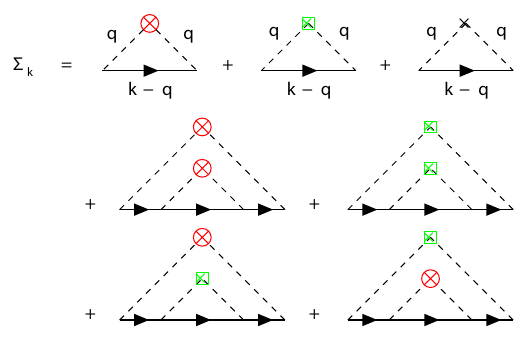}
\caption{Self-energy\cite{RevModPhys.57.287} under Born approximation. Red (circle), green (box) and black nodes represent $AB$-type, $BA$-type and $AA$-type impurity scatter vertex, respectively. For tidiness, we omit the momenta of Green functions in 2-loop diagrams.}\label{fg4}
\end{figure}

On this platform, we perform disorder average to obtain the self-energy $\Sigma_{\vec{k}}$ under Born approximation. When calculating the self energy, we only consider terms whose values are much larger or at least comparable with the next-nearest-hopping. Given that $U_{\vec{k}\vec{k^{\prime}}}^{AA}$ is about one order smaller than other two $U$s, 1-loop diagram constructed by $AA$-type vertex has the same order as 2-loop diagrams that do not include $AA$-type vertex. Thus, only those Feynman diagrams showed in FIG. \ref{fg4} are included in our calculation of self-energy. The choice of coefficients is the same as above. We set the strength and density of the impurities on $AA$ regions, $AB$ regions and $BA$ regions to be equal.

\begin{figure}[htbp]
\centering
\includegraphics[width=8cm]{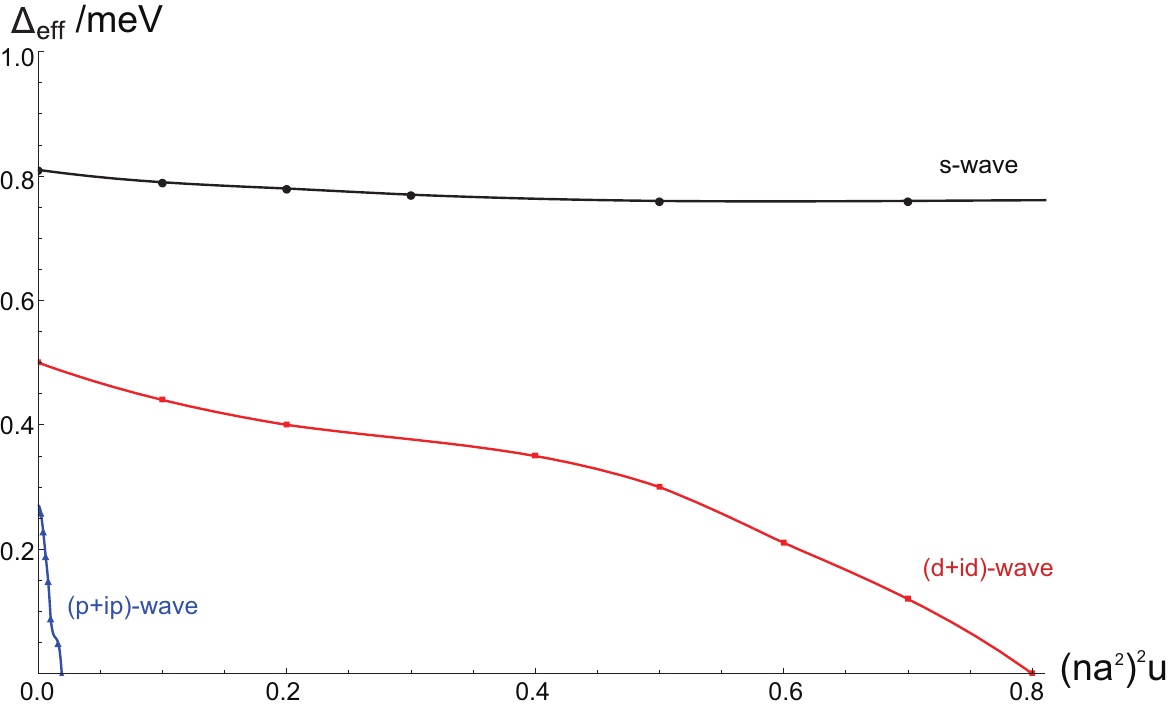}
\caption{Phase diagrams. From this figure we can see that enough strength or concentration of impurities will destroy $(d+id)$-wave and $(p+ip)$-wave superconductivity, while the superconducting gap of $s$-wave phases can remain finite.}\label{fg3}
\end{figure}

A result that can be obtained from disorder average is the phase diagram, which reflects how the effective superconducting gap relies on the density and strength of impurities, $(na^2)^2 u$. Keeping other coefficients invariant, we vary $(na^2)^2 u$ from 0.0 to 1.0 and find the corresponding value of the effective gap. Results are shown in FIG. \ref{fg3}. According to the phase diagrams, effects of impurities in different superconductivity phases are different. In $(d+id)$-wave phase and $(p+ip)$-wave phase, strong or dense impurities will destroy the superconductivity while in $s$-wave phase they will not.

\subsection{\label{sec:level2}Explanation for an Anomalous Feature of Some Figures}

Some figures of local DOS in FIG. 8 and FIG. 9 show an anomalous feature, that for $(d+id)$-wave phase and $(p+ip)$-wave phase, the local DOS of two gap edges seemingly lose particle-hole symmetry in strength. Indeed, since under our choice of coefficients, the value of superconductor gap is comparable with Bogoliubov band gap at $K$ point in the Brillouin zone, as shown in FIG. 13. However, particle-hole symmetry of the strength of the local DOS of two gap edges only occurs when the value of superconducting gap is much smaller than that of band gaps. Therefore, nothing will guarantee the particle-hole symmetry of the strength of the two gap edges in the DOS of $(d+id)$-wave phase and $(p+ip)$-wave phase in our model.

\begin{figure}[htbp]
\centering
\includegraphics[width=8cm]{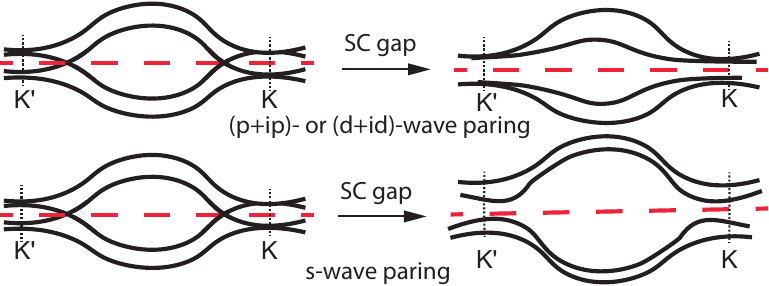}
\caption{Bogoliubov bands with and without superconductivity. In our choice of coefficients, because of the form factor of $(p+ip)$-wave and $(d+id)$-wave paring, the superconducting gap is comparable with the band gap at $K$ point, as illustrated in the upper half of FIG. 13. For $s$-wave pairing, the superconducting gap is always much smaller than the band gap at $K$ point as illustrated in the lower half of FIG. 13. Therefore, for $s$-wave phase, the local DOS of two gap edges have particle-hole symmetry in strength.}
\end{figure}

\bibliographystyle{apsrev4-1}

\bibliography{TBGImp.bib}
\end{document}